\documentclass[twocolumn]{aastex631}

\usepackage{url}
\usepackage{amsmath}
\usepackage{amssymb}
\usepackage{graphicx}
\usepackage[version=4]{mhchem}
\usepackage{xcolor}

\shorttitle{Radiative transfer modeling of Haumea's dust ring}
\shortauthors{Kalup, Moln\'ar \& Kiss}

\graphicspath{{./}{figures/}}

\begin{document}

\title{Radiative transfer modeling of Haumea's dust ring}

\author[0000-0002-1663-0707]{Csilla Kalup}
\affiliation{Konkoly Observatory, HUN-REN Research Centre for Astronomy and Earth Sciences, Konkoly Thege 15-17, H-1121~Budapest, Hungary}
\affiliation{CSFK, MTA Centre of Excellence, Budapest, Konkoly Thege 15-17, H-1121, Hungary}
\affiliation{ELTE E\"otv\"os Lor\'and University, Institute of Physics and Astronomy, P\'azm\'any P. st. 1/A, 1171 Budapest, Hungary}

\author[0000-0002-8159-1599]{L\'aszl\'o Moln\'ar}
\affiliation{Konkoly Observatory, HUN-REN Research Centre for Astronomy and Earth Sciences, Konkoly Thege 15-17, H-1121~Budapest, Hungary}
\affiliation{CSFK, MTA Centre of Excellence, Budapest, Konkoly Thege 15-17, H-1121, Hungary}
\affiliation{ELTE E\"otv\"os Lor\'and University, Institute of Physics and Astronomy, P\'azm\'any P. st. 1/A, 1171 Budapest, Hungary}

\author[0000-0002-8722-6875]{Csaba~Kiss}
\affiliation{Konkoly Observatory, HUN-REN Research Centre for Astronomy and Earth Sciences, Konkoly Thege 15-17, H-1121~Budapest, Hungary}
\affiliation{CSFK, MTA Centre of Excellence, Budapest, Konkoly Thege 15-17, H-1121, Hungary}
\affiliation{ELTE E\"otv\"os Lor\'and University, Institute of Physics and Astronomy, P\'azm\'any P. st. 1/A, 1171 Budapest, Hungary}

\begin{abstract}
Among the growing number of small body rings in the solar system, the ring of Haumea has a special status as it is likely suitable for direct imaging in the visible and submillimeter wavelengths. In this paper, we highlight its sole detectability among Centaur/TNO rings using both the ALMA and the James Webb Space Telescope to provide direct constraints on the ring composition for the first time. To overcome the limitations of the currently used simple ring models, we introduce radiative transfer modeling for small body ring systems. Here we perform a thorough analysis of the Haumea ring considering different materials and grain sizes, assuming that the ring consists of small particles with sizes below 1\,mm. We present spectral energy distributions of each model for future comparison with multiwavelength measurements, providing a diagnostic tool to determine the dominant grain size and characteristic material of the ring, which are essential inputs for ring formation and evolution theories. Our results also show that for some sub-micron carbon-like or silicate grains, their mid-infrared excess can be detected even if the ring is not resolved, providing a tracer for small grains around the object.
\end{abstract}
\keywords{Solar system (1528) -- Dwarf planets (419) --Planetary rings (1254)}


\section{Introduction}

In 2013, the astonishing discovery of the rings around the Centaur-type object Chariklo \citep{Braga-Ribas2014} unveiled the first small body in the Solar System hosting a ring system, beyond the giant planets. This detection directed focus onto another Centaur, Chiron, offering a possibly similar ring system interpretation of reanalyzed, previous stellar occultation measurements \citep{Ortiz2015, Ortiz2023}. On 2017 January 21, multichord stellar occultation observations of Haumea revealed a $\sim$70\,km-wide ring with a radius of $2287\pm80$\,km \citep{Ortiz2017}. This finding was a big surprise, as while Centaurs are a dynamically rather unstable class of small bodies with orbital semi-major axes confined to between those of Jupiter and Neptune, and are capable of showing both asteroid and comet characteristics, the more distant Trans-Neptunian Objects (TNOs) are located in a completely different dynamical environment, leading to further questions of how a small body ring system can be formed. Recently, multiple rings were discovered around the large Kuiper belt object Quaoar, too \citep{Morgado2023,Pereira2023}. 

Haumea is an exotic, very elongated and fast rotating (P\,=\,3.9\,h) dwarf planet among the icy debris of the Kuiper belt \citep{Brown2005,Rabinowitz2006}. It is the largest member of the Haumea dynamical family, the only known collisional group among TNOs so far \citep{Brown2007}. All the family members, along with its two known moons, Hi'iaka and Namaka, show a high albedo, related to nearly pure crystalline water-ice surfaces \citep{Barkume2006,Trujillo2007,Pinilla-Alonso2009}. These properties point to a formation scenario via a graze-and-merge type of collision \citep{Proudfoot2022}. Alternative theories about the formation, e.g., by \cite{Noviello2022} assume a more complex sequence of events: a giant impact followed by internal chemical evolution, hydration of minerals, and ejection of material due to high-speed rotation.

Based on the occultation, the ring's dimensions, orientation, opacity and estimated reflectivity have been obtained \citep{Ortiz2017}. It is known that the ring is co-planar with the orbit of Hi'iaka, the major satellite of Haumea, and it is presumably in a 3:1 spin-orbit resonance with Haumea. The resonance may have a stabilizing effect within the rapidly changing gravitational potential of the triaxially-shaped dwarf planet \citep{Sicardy2020book}. The discovery generated a large number of studies on the ring's dynamics \citep[see e.g.][]{Kovacs2018,Sicardy2019,Sicardy2020,Sanchez2020,Kondratyev2020,Marzari2020,Ikeya2024}.
However, little is known about the ring itself, even though that would be a crucial step to derive clues to its origin, evolution and connection to Haumea and its dynamical group. Occultation chords define the radius of the ring, r\,=\,2287$_{-45}^{+75}$\,km, and using an apparent opacity of p'\,=\,0.555 (or an equivalent apparent optical depth of $\tau'$\,=\,$-\ln{(1-p')}$) a ring width of w\,=\,70\,km can be calculated, as discussed by \citet{Ortiz2017}. They also estimated the contribution of the ring to the total visible range of brightness using absolute brightness measurements of Haumea in 2005 and 2017, and considering the ring geometry (pole solution/opening angle) obtained from the occultation measurements. Their analysis excludes a high reflectivity ring (I/F\,$\gtrsim$\,0.36), and favors a ring with I/F\,=\,0.09, which corresponds to a contribution of $\sim$2.5\% to the total brightness of the system in 2017. Their obtained reflectivity of I/F\,=\,0.09 is comparable to Chariklo's main ring \citep{Braga-Ribas2014,Leiva2017}, and brighter than the rings of Uranus \citep[I/F\,$\approx$\,0.05,][]{Karkoschka2001}, but dimmer than the rings of Saturn \citep[I/F\,$\approx$\,0.5,][]{French2007}. 

This indirect estimate on the ring's reflectivity is, however, in contradiction with the canonical ring formation theory, which states that the rings are formed from the material of shattered satellites. If the ring was formed from the material of such a bright and water-rich icy satellite, which is typical of the Haumea family, its reflectance should be closer to that of the Saturnian rings (I/F\,$\approx$\,0.5). Darker ring material indicates a different formation mechanism, e.g., a recent disruption of a lower albedo object. Collisional simulations indicate that disks may form in a significant fraction of giant impacts -- similar to the one that created the present-day Haumea and presumably its dynamical family \citep[e.g.][]{Arakawa2019}, but the origin of the ring from these events also raise the question of how the ring could remain stable for such a long time. Since the gravitational field of the elongated Haumea and the 3:1 resonance may have a stabilizing effect on the ring \citep{Sicardy2020book}, it may also have prevented the ring material from forming a satellite. In addition to collisional origins, rings may form by other types of mass loss, such as intermittent releases from the surface \citep{Noll2023}.

In this paper, we investigate the detectability of all known Centaur/TNO ring systems and show Haumea's vitally important role among them. Then, to go beyond the capabilities of the currently used simple ring models, we present radiative transfer modeling for a ring system of a small body for the first time in the literature, which enables a more detailed treatment of the dust grain properties of a ring. While stellar occultations are predominantly measured in the visible light range, observations in the near-infrared and submillimeter regimes are also essential to derive information on the ring properties. 
Currently we lack any clues on the composition, i.e. material type or grain size distribution, for Haumea's ring, or for any other small body ring system. Based on dynamical considerations and examples from giant planet ring systems, rings can be made of large boulders. However, small body ring systems are in completely different dynamical environments and their study requires other processes to be considered.    
Here, we simulated rings using different materials, assuming particles below 1\,mm grain sizes, which can also be seen in some Saturnian and Uranian rings \citep{Hedman2018}. Due to the optical properties of these kinds of grains, they show much more pronounced spectral attributes compared to large particles \citep{BH87}. The aim of our work is to provide models for future, multiwavelength observations, where the presence or lack of the wavelength dependence of the observations can support or exclude the existence of small grains in the Haumea ring system, revealing similarities or differences to planetary rings or to comet-like compositions. This analysis not only opens up ways to further studies on the possible mixture of large boulders and small particles, but can also serve as a case study for future analysis of small body ring compositions, such as that of the recently detected mid-infrared excess of the dwarf planet Makemake, where one of the possible explanations is a yet undetected ring containing sub-micrometer-sized carbonaceous dust grains \citep{Kiss2024b}. 



In Sect.~\ref{sect:simpleringmodel} we present the simple ring model, previously applied to estimate the visible brightness and thermal emission of different Centaur and dwarf planet ring systems, and use it to examine wether the presently known Centaur/TNO ring systems could be resolved with available instruments; in Sect.~\ref{sect:comp} we elaborate on the possible ring scenarios based on analogues of planetary rings and comets; in Sect.~\ref{sect:radmodel} we summarize our radiative transfer model and its input parameters; in Sect.~\ref{sect:results} we present our results; and our summary and conclusions are given in Sect.~\ref{sect:summary}.

\section{A Simple Ring Model}
\label{sect:simpleringmodel}

Our current knowledge on the ring is limited to the properties derived from the occultation data and the brightness estimates of the system at different epochs (observing geometries) in the visible and thermal infrared ranges. \cite{Muller2019} analyzed the thermal emission of the Haumea system using a simplified version of Saturnian ring models. A detailed description of this simple model and the related references can be found in \citet{Lellouch2017}, where it was used to estimate the brightness of Chariklo in infrared and submillimeter wavelengths; we just recite the main equations here. 
According to this model, the temperature of the ring particles can be obtained as:
\begin{equation}
    T_p = \Bigg( \frac{(1-A_r) \sin{B'} (1-e^{-\tau/\sin B'}) S_\odot}
    {(1-e^{-\tau})\epsilon_{r,b} \, \sigma \, f \big(1 - \frac{6(1-e^{-\tau})}{4\pi} \big) r_h^2}  \Bigg)^{1/4}
\end{equation}
where $A_r$ is the Bond albedo, $B'$ is the solar elevation above the ring plane,  $\tau$ is the optical depth, $S_\odot$ is the solar constant, 
$\epsilon_{r,b}$ is the bolometric emissivity of the ring particles, $\sigma$ is the Stefan-Boltzmann constant, $f$ is the rotation rate factor (assumed to be f\,=\,2 here) and $r_h$ is the heliocentric distance. The Bond albedo can be described as the product of the reflectivity and the phase integral of the ring. The brightness temperatures can be obtained as:
\begin{equation}
    B_\nu (T_B(\lambda)) = \epsilon_{r,\lambda} (1-e^{-\tau/\sin B})B_\nu(T_p)
\end{equation}
where $B_\nu (T)$ is the Planck function, $\epsilon_{r,\lambda}$ is the spectral emissivity, $B$ is the elevation of the observer above the ring plane, and it is assumed that  $\epsilon_{r,\lambda}$\,=\,$\epsilon_{r,b}$\,=\,1.

\begin{figure}[ht!]
\begin{center}
\includegraphics[width=\columnwidth]{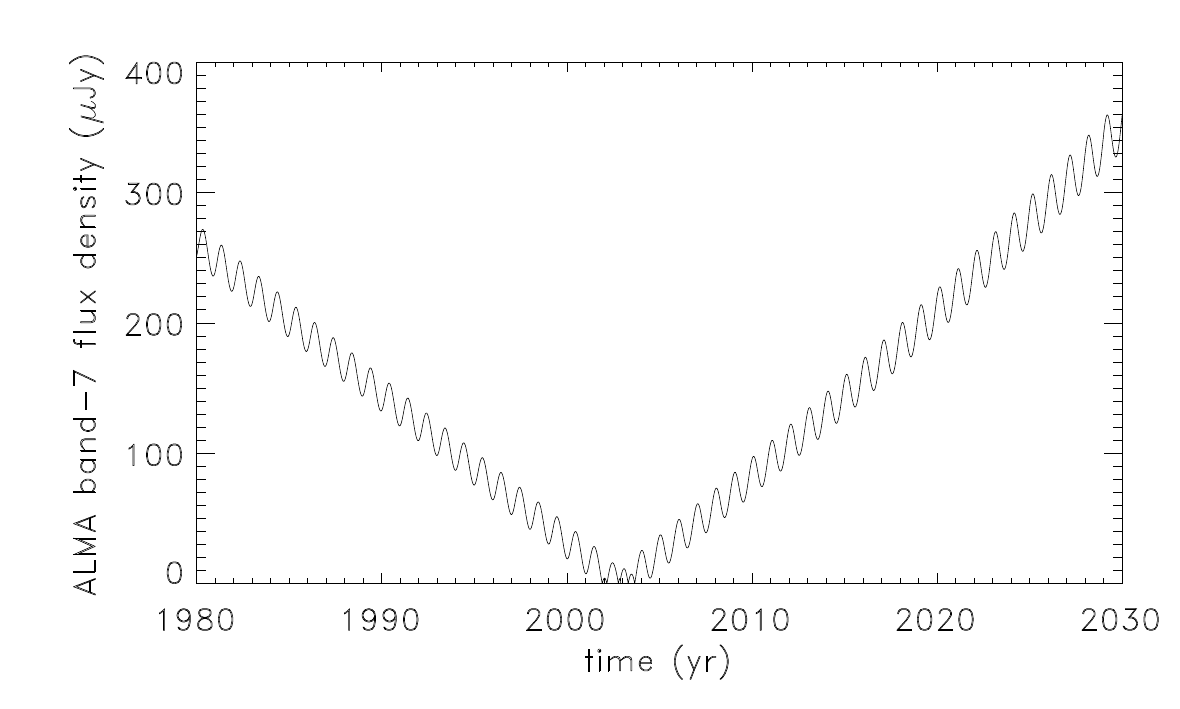}
\caption{ALMA band-7 flux density prediction of the simple ring model assuming apparent optical depth of $\tau\,=\,0.81$, ring radius of r\,=\,2287\,km and width of w\,=\,70\,km, reflectivity of I/F\,=\,0.09 and phase integral of q\,=\,0.26. The minimum around 2003 is due to an edge-on observing geometry of the ring, while the small scale variation is due to the orbit of the Earth around the Sun. Reproduced from \citet{Kalup2023}.
\label{fig:almaband7}}
\end{center}
\end{figure}

Based on this model, \cite{Muller2019} estimated the contribution of the ring to the thermal emission of the system between 1980 and 2030 in the \textit{Herschel} PACS and SPIRE bands, as well as derived upper limits on the total thermal brightness of $2.52\cdot 10^{-3}$\,Jy and $2.66\cdot 10^{-3}$\,Jy at 100\,$\mu$m and 160\,$\mu$m, respectively. Applying the same model, \citet{Kalup2023} partly revisited the thermal model flux prediction and provided flux density estimates for the ring in ALMA band-7 measurements using various reflectivities and updated opacity and phase integral values. Based on these calculations, we provide a reproduction in Figure $\ref{fig:almaband7}$, and derive $F_{\mathrm{ALMA}}$ flux densities at $\sim870\,\mu$m for 2017 January 21 and at the beginning of 2025 for comparison with the radiative transfer approach. We found that $F_{\mathrm{ALMA}}^{2017}=181\,\mu$Jy and $F_{\mathrm{ALMA}}^{2025}=287\,\mu$Jy, assuming this simple ring model.

\subsection{Direct Detectability of Centaur/TNO Rings}

Direct imaging of a ring requires that the ring is well separated from the primary body (i.e. the spatial resolution is good enough with the actual instrument), and bright enough to be detected close to the much brighter primary. We examined the ring detectability for the known ring systems around Chariklo, Quaoar and Haumea using the simple ring models. Table~\ref{table:ringsystems} shows the main known and estimated properties of the Chariklo, Quaoar and Haumea ring systems.

The required spatial resolution to resolve the rings is in the $\leq$30\,mas range (see Table~\ref{table:ringsystems}). Currently, only the Atacama Large Millimetre Array radio telescope system (ALMA) and the James Webb Space Telescope (JWST) are capable to achieve it. While adaptive optics (AO) systems of large telescopes (e.g. VLT or Keck, typically working in the near- and mid-infrared) are theoretically capable of achieving the required resolution, these would require extremely long integration times, exceptional PSF stability and suitable, nearby reference stars to allow for an efficient PSF subtraction to detect the very faint and close-by ring. Studies with AO systems have been performed on dwarf planet satellites \citep[see e.g.][]{Brown2007,Fraser2013} but the vast majority of the TNO binary detections and characterisations were performed with the Hubble Space Telescope to avoid the technical difficulties of AO systems. In those cases, the satellites were at significantly larger distances than the extent of the Centaur/TNO rings, therefore detection of the Centaur/TNO rings would be even more challenging with these AO instruments. With the ALMA system, there are ongoing measurements on the Haumea system in band-7 ($\sim$870\,$\mu$m) with the Cycle-9 antenna configuration (proposal ID 2022.1.01753.S)\footnote{https://almascience.eso.org/observing/highest-priority-projects} in the thermal emission regime, which configuration allows the $\leq$30 mas resolution. In the visible wavelength range, the best detectability could be achieved with in the shortest wavelength F070W band of the NIRCam instrument, which provides a spatial resolution of 29\,mas, while also having a high sensitivity and good PSF stability. This narrow PSF is essential to separate the light from the body of Haumea and that of the ring. We also note that due to the high contrast between Haumea and its ring, the ring emission is only obtainable as intensity enhancements on both sides of Haumea after creating differential images from the combined PSF of the Haumea-ring system convolved with the shape of Hauema’s body.

\begin{table*}[t!]
    \centering
    \hspace{-2.5cm}
    \begin{tabular}{cccccccccc}
    \hline
         System / ring & r & w & r' & $\tau_N$ & F$_{tot}^{\mathrm{ALMA}}$ & F$_{tot}^{\mathrm{JWST}}$ & B$_{\mathrm{ALMA}}$ & B$_{\mathrm{JWST}}$ & References\\

        & (km) & (km) & mas &  & ($\mu$Jy) & ($\mu$Jy) & ($\mu$Jy/beam) & ($\mu$Jy/beam)\\
        
    \hline
         Chariklo/C1R & 386 & 6.9 & 31 & 0.2 & 102 & 23 & n.r. & n.r. & B14, L17, M21 \\
    \hline
         Quaoar/Q1R/max & 4057 & 52 & 128 & 0.04 & 186 & 17 & $\sim$16 & $\sim$0.5 & M23, P23, K24\\
         Quaoar/Q1R/mean & 4057 & -- & -- & -- & 93 & 5 & $\sim$8 & $\sim$0.17 & K24\\
    \hline
         Haumea/HR1 & 2287 & 70 & 63 & 0.36 & 287 & 18 & $\sim$24 & $\sim$1.5 & O17, K23 \\
    \hline
         
    \end{tabular}
    \caption{Main characteristics of the known Centaur/TNO ring systems calculated for at the beginning of 2025. In the case of multiple ring systems the main ring is considered. The columns are:
    object/ring name; 
    ring radius; ring width; apparent ring radius for 'terrestial' observer; normalized optical depth; 
    estimated visible range total flux density, using the same method as in \citet{Kalup2023}; estimated brightness (n.r. $\equiv$ not resolved); and references to the ring properties: B14 -- \citet{Braga-Ribas2014}, L17 -- \citet{Leiva2017}, M21 -- \citet{Morgado2021}, M23 -- \citet{Morgado2023}, P23 -- \citet{Pereira2023}, O17 -- \citet{Ortiz2017} and references for already existed total flux predictions: K23 -- \citet{Kalup2023}, K24 -- \citet{Kiss2024}. For the Quaoar Q1R ring a 'maximum' and a 'mean' scenario was also calculated (see the text for details, and also in \citet{Kiss2024}).}
    \label{table:ringsystems}
\end{table*}

Due to insufficient spatial resolution, Chariklo's rings can't be detected separately from the main body. Quaoar's main ring (Q1R) is notably larger than the rings of Chariklo and Haumea, however, as the ring is an extended feature, the calculated total flux densities correspond to 'per beam' brightnesses, resulting in much fainter rings for the observations, even when considering the densest part as representative for the whole ring ('maximum' case in Table~\ref{table:ringsystems}). We also list a more feasible 'mean' Q1R ring value, where we assume that the ingress and egress occultation measurements are representative of the occurence of the respective ring section with different width and optical depth \citep{Kiss2024}. Among the known Centaur/TNO rings, only the Haumea ring is bright enough to be detected by both ALMA/Band-7/C9 and JWST/NIRCam/F070W with reasonable integration times, making it one of the most important touchstones not only for Kuiper belt, but also for small body ring formation and evolution theories.

\section{Comparison with Planetary Rings and Comets}
\label{sect:comp}

In our solar system, Saturn and Uranus harbor the most well-characterized rings, serving as natural references for studying small body ring systems. As a result of angular momentum conservation and dissipative interparticle collisions, planetary rings form exceptionally flat disks with at characteristic thickness between 5--50 meters, on the order of a few times of the diameter of the largest particles \citep{Nicholson2018}. For ring particles in the size range of cm to few meters, a power-law of $\sim$$r^{-q}$ describes the grain size distribution, where the exponent is constrained to the interval of $2.75\leq q \leq 3.5$. For larger sizes, the distribution has a steep cutoff. Detailed models of kinetic processes with a dynamic balance between aggregation and fragmentation show that this power-law distribution with a large size-cutoff seems universal, and it is expected for any ring system where collisions play a role \citep{Brilliantov2015}. The minimum particle size is on the order of millimeters, but the amount of these sub-centimeter particles is lower than expected from the extrapolation of the power-law distribution \citep{Harbinson2013}, which can be explained by the cohesive forces between particles, causing them to stick to the surface of larger ones \citep{Ohtsuki2020}. In the case of Uranus, several kinds of methods strongly suggest that its ring particles are similarly large, with a minimum radius of several centimeters, and a maximum of few meters. One exception is the $\lambda$ ring, which is believed to consist of much smaller, micron-sized particles \citep{Nicholson2018}. The Phoebe ring of Saturn is also assumed to be dominated by small grains, however, in this latter case the actual grain size is not well constrained \citep{Verbiscer, Hamilton}. 

Regarding the composition of these rings, Saturn and Uranus show completely different characteristics. For Saturn, the ring particles are composed of more than 90\%--95\% crystalline water ice with a small addition of silicates \citep{Cuzzi2010}. Compared to its icy satellites, the ring looks generally much redder at visual wavelengths, which can be interpreted as the rings containing much smaller, embedded or intramixed grains of organic material (tholins) and amorphous carbon, too \citep{Poulet2002, Poulet2003}. For Uranus, the observed dark and spectrally featureless data showing very low reflectivity and flat spectra indicates that a carbon-rich material is most likely the dominant component \citep{Nicholson2018}.

Centaur/TNO objects are not only much smaller than giant planets, but also possess unique locations at special dynamical environments and characteristics (e.g., Centaurs exhibit both asteroid- and comet-like signatures). Moreover, a recent study by \cite{Kiss2024b} revealed an unexpected mid-infrared excess from the dwarf planet Makemake, which cannot be explained by the thermal emission of solid bodies irradiated only by the Sun at that heliocentric distance. However, one of the proposed explanations that can reproduce the observation is a ring around Makemake made of very small carbonaceous grains. Besides this, another recent occultation measurement of a small object in the Solar System also indicates that very small grains may dominate some of these ring systems (P. Santoz-Sanz, private communication). For small grains, submicron cometary dust can be used as an example; that tends to be dominated by amorphous carbon, while the submicron silicate mass in it tends to contain primarily amorphous refractory grains, mainly Mg-rich olivine and pyroxene \citep{Wooden2017,Harker2023}.

We created ring models with different compositions using various material and grain size combinations, inspired by planetary rings and comets, for the 2017 observing geometry of the Haumea ring system. We calculated carbon, graphite and water ice models, as well as olivine and pyroxene with varying magnesium content, and covered grain sizes from 0.1 to 1000\,micron for each material. Although planetary rings exhibit a power law distribution of grain sizes made out of a mixture of materials, in this analysis we focus on identifying the potentially dominant component and particle radius for the Haumea ring. We choose 1\,mm as the upper limit of the grain sizes, because below this radius the observed wavelength dependence of the ring's SED can give indications on the possible constitution of the ring’s dust particles \citep{BH87}. Particles larger than 1\,mm do not show significant differences at the visible, infrared and submillimeter regimes for any given material. 

\section{Radiative Transfer Modeling}
\label{sect:radmodel}

The radiative transfer modeling of the dust grains in the ring is performed using the RADMC-3D \citep{radmc} freely available, open source software package for diagnostic radiative transfer calculations in arbitrary 1D, 2D or 3D geometries. For a given geometrical distribution of gas and/or dust, it calculates what the images and/or spectra of the setup look like when viewed from a certain position including optical depth calculations, both in scattered light and in thermal emission. 

\subsection{General Setup}

We created a coordinate system, whose center coincides with the center of the simulation volume, and is fixed to the ring in a way that the ring's symmetry plane is the XY-plane. The Sun is located at a distance of $r_h$ from the center of the coordinate system on the XZ-plane, with $x_\sun$ and $z_\sun$ coordinates corresponding to the actual inclination of the ring, outside the simulation volume. The Sun is considered as a black body with $T_{\rm eff} = 5780$\,K temperature and $R_\odot = 6.96\times10^5$\,km radius. 

The materials used for radiative transfer modeling are characterized by their composition, porosity and size distribution. We use the {\sl optool}  package \citep{optool} to obtain the complex  opacities of the grains, using materials from {\sl optool}'s material library, assuming specific grain sizes. We consider different types of materials in different simulations, but we assume that the whole simulation volume is homogeneous material-wise. Anisotropic scattering is considered, and treated by applying the Henyey-Greenstein function (option {\sl scattering\_mode\_max\,=\,2} in RADMC-3D), and using the scattering opacity and $g$ anisotropy parameters obtained for the specific material with {\sl optool}. As a first approximation, we ignored porosity in our models. However, based on our preliminary tests, changing the porosity for larger grains changes only the scattered light range significantly: the more porous the grain, the fainter the emission.

\subsection{Ring Element Model}

We approximate the ring by modeling the radiative transfer in a small element of the ring, and then multiplying the flux densities by the ratio of the ring area ($A_r$) to the area of the simulated region ($A_s$).  The apparent area of the ring is: 
\begin{equation}
    A_r = \pi( (r_i+w)^2 - r_i^2 )\sin(B)
\end{equation}
where $r_i$ and $w$ are the inner radius and width of the ring, and $B$ is the opening angle. 
This model implicitly assumes that the ring is homogeneous, thin (like in the simple `analog' ring model in Sect.~\ref{sect:simpleringmodel}), and that edge effects are ignored. 
In practice, we used a 128$\times$128$\times$8 cell simulation volume, which is illuminated under the proper solar geometry with respect to the ring, and seen under the proper geometry from Earth in 2017. The simulation volume is filled homogeneously with the actual material chosen. The apparent, projected image of the simulated ring element is not homogeneous due to projection effects, and has lower optical depths towards the edges. However, it has the correct values in its central part, which corresponds to the surface brightness values as if it was seen in a homogeneous, full ring. We used those projected pixels for which $\tau$\,=\,max($\tau$) in the visible-range \textit{R}-band, also calculated by RADMC-3D. The mean values of these selected pixels provide the flux density at a specific wavelength. The area of the simulated element $A_s$ is simply the (apparent) area of a projected pixel. 

\subsection{Differences with Respect to the Simple Ring Model}
\label{sect:radtr}

\begin{figure}[ht!]
    \centering
    \includegraphics[width=\columnwidth]{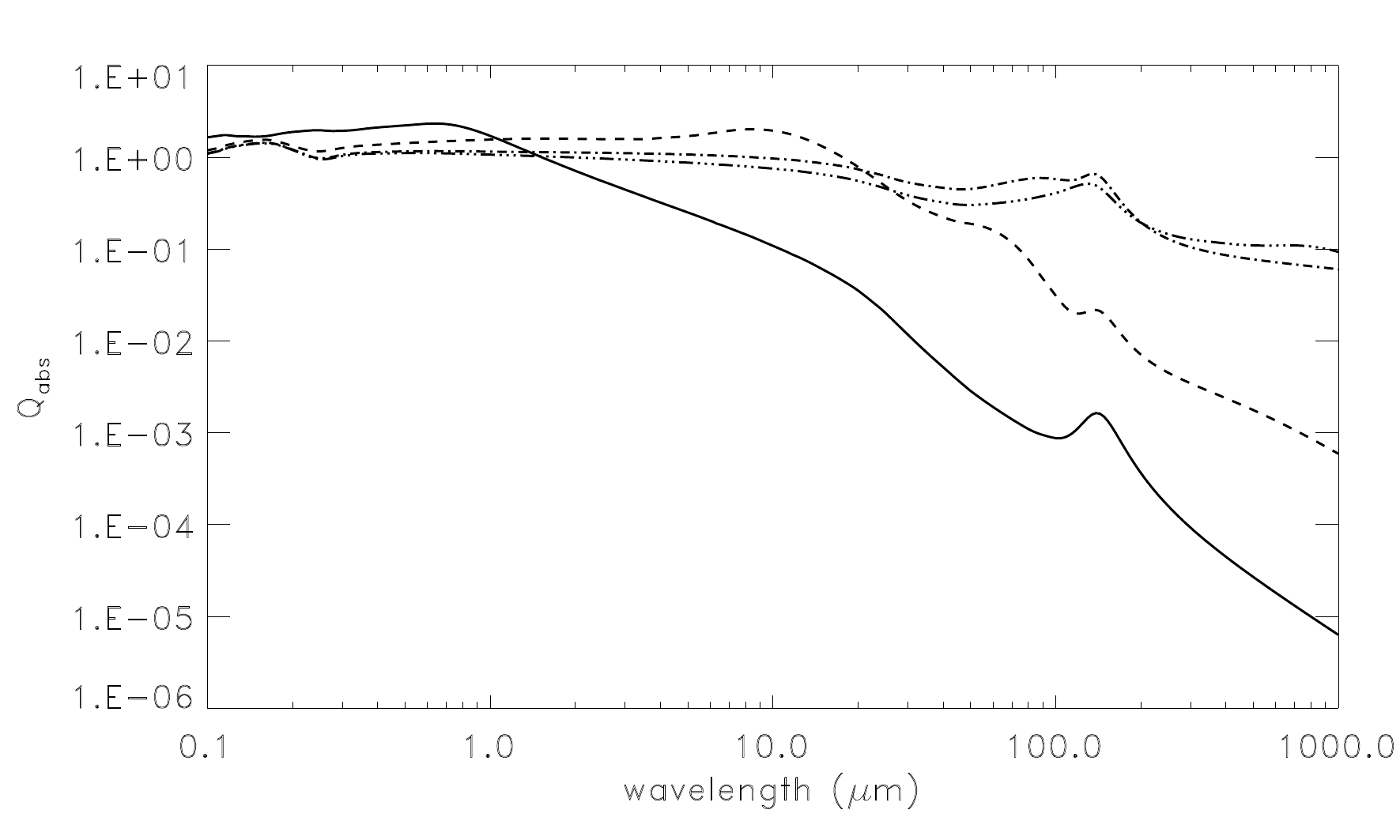}
    \includegraphics[width=\columnwidth]{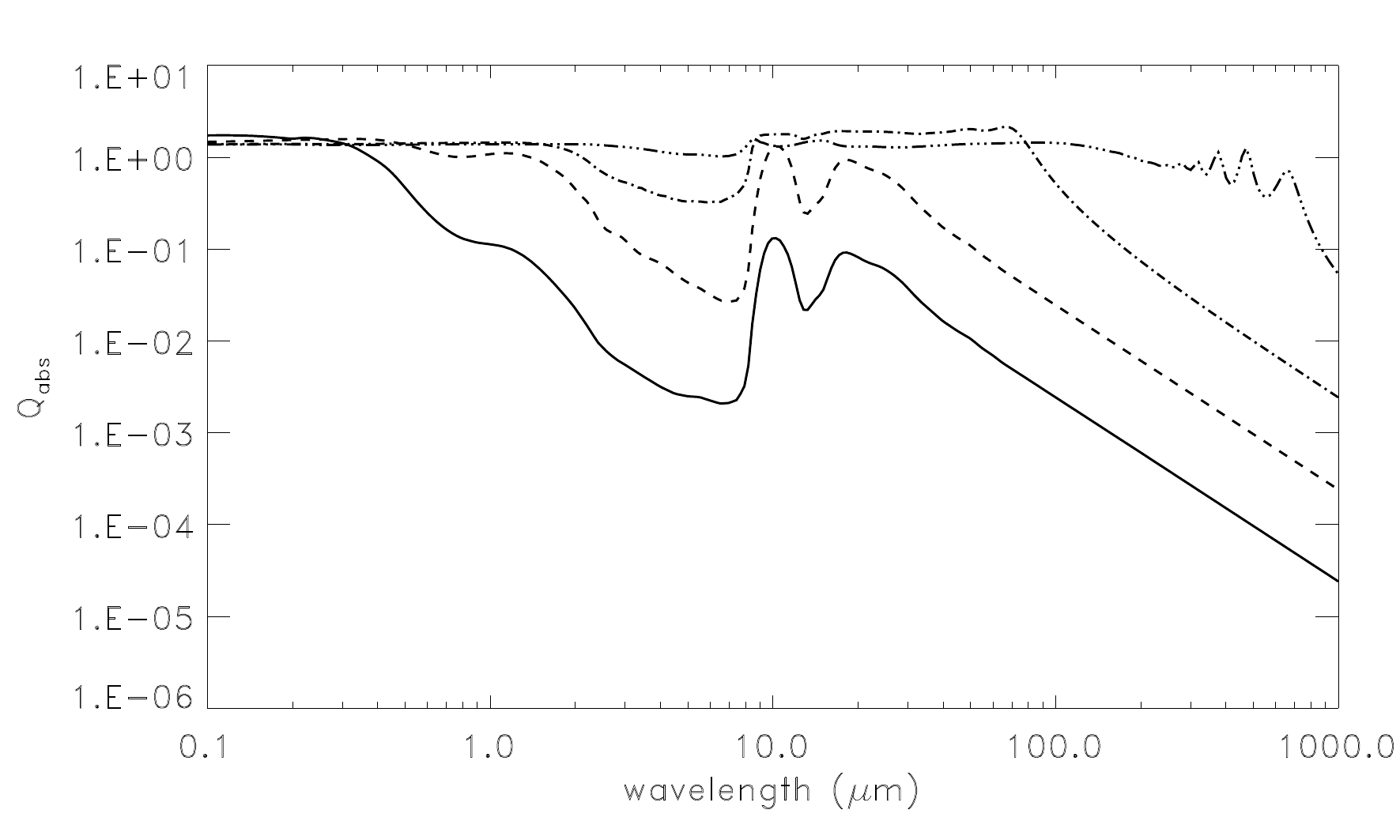}
    \caption{Absorption efficiency for graphite (top panel) and olivine grains with 50\% Mg-content (bottom panel). The curves correspond to different grain sizes. Solid curve: 100\,nm, dashed curve: 1\,$\mu$m, dash-dotted curve: 10\,$\mu$m, dash-tripple-dotted curve: 100\,$\mu$m.}
    \label{fig:qabs}
\end{figure}

There are two main features in the radiative transfer models that may result in significant deviations from the predictions of the simple ring emission model. First, in addition to the compositional differences, the emissivity of the grains is strongly wavelength and grain size dependent \citep[see e.g.][]{DL84}. The spectral emissivity equals to the absorption efficiency at a specific wavelength \citep[see e.g.][]{BH87} which can be calculated as $Q_{abs}(\lambda) = (4/3)a\rho\kappa_{abs}$, where $a$ is the grain size, $\rho$ is the density of the grain, and $\kappa_{abs}$ is the mass absorption coefficient (obtained from {\sl optool} in our case). To demonstrate this effect, we plotted $Q_{abs}$ for graphite and olivine grains using different grain sizes in Fig.~\ref{fig:qabs}. As the figure shows, the emissivity typically decreases for longer wavelengths, especially in the far-infrared--submm range, and larger grains have higher emissivity, also leading to higher temperatures of smaller grains. 

\begin{figure}[ht!]
    \centering
    \includegraphics[width=\columnwidth]{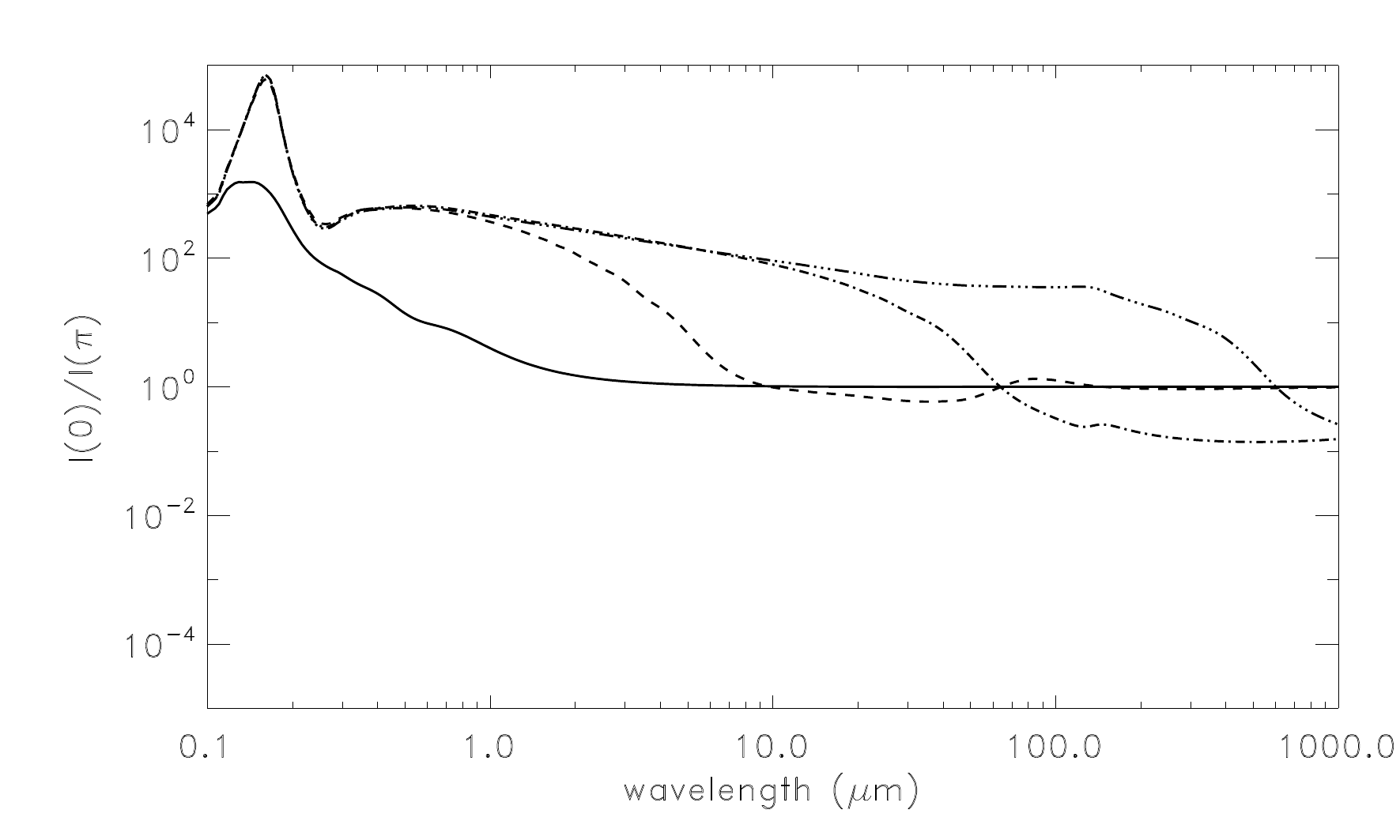}
    \includegraphics[width=\columnwidth]{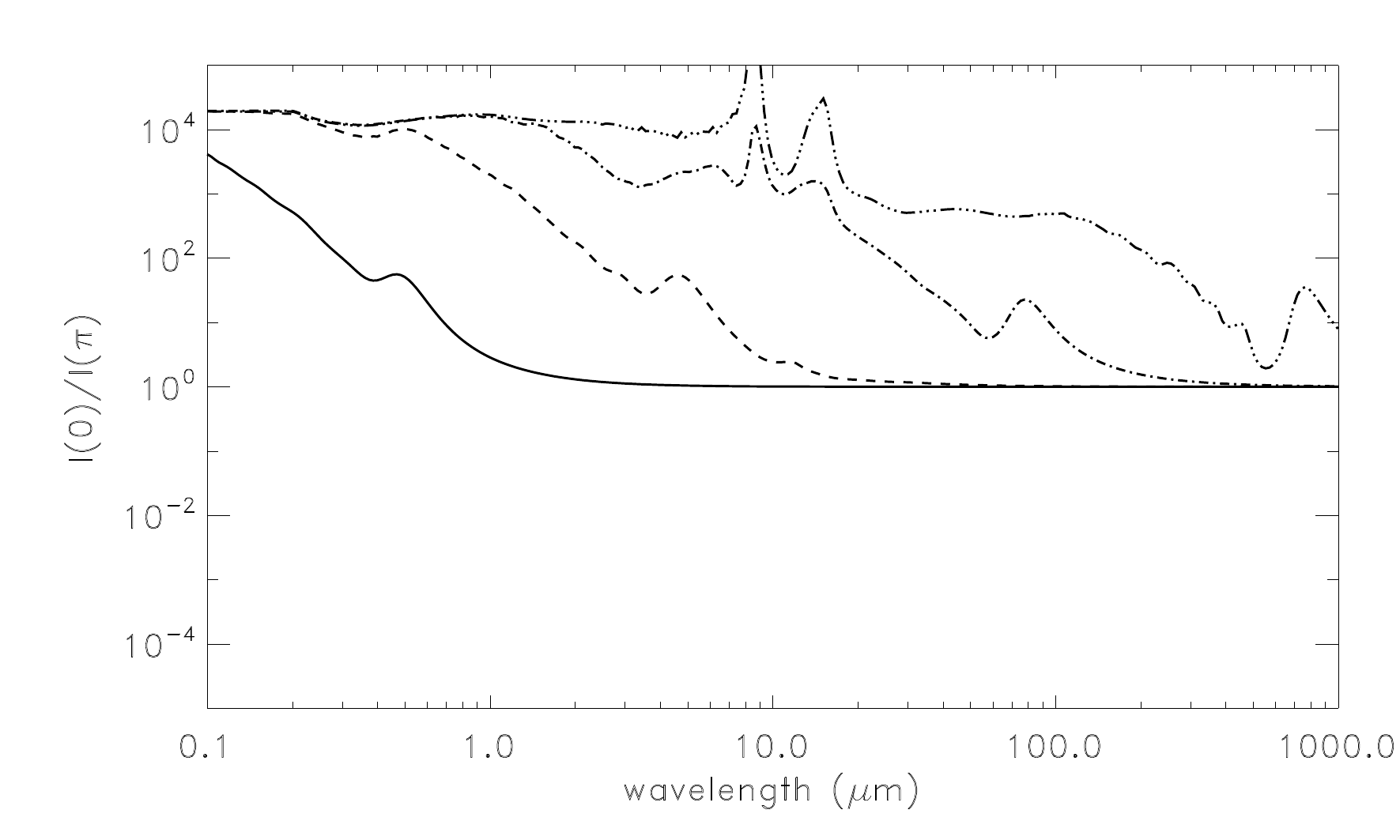}    
    \caption{Wavelength dependence of the ratio of the forward- and backward-scattered radiation, $I(0)/I(\pi)$, for graphite (top panel) and olivine grains with 50\% Mg-content (bottom panel). The curves correspond to different grain sizes. Solid curve: 100\,nm, dashed curve: 1\,$\mu$m, dash-dotted curve: 10\,$\mu$m, dash-tripple-dotted curve: 100\,$\mu$m.}
    \label{fig:g}
\end{figure}

Another important feature is the consideration of anisotropic scattering. Here in our model, as mentioned in Sect.~\ref{sect:radtr}, this is handled using the Henyey-Greenstein function and the $g$ anisotropy coefficient. In Fig.~\ref{fig:g} we plot the ratio of the forward- and backward-scattered intensity, $ I(0)/I(\pi) = [(1+g)/(1-g)]^3$,  as a function of wavelength for the two example materials and for different grain sizes. Essentially, for all grain types and sizes we use in our study here, the grains are very strong forward scatterers ($I(0)/I(\pi)$\,$\gg$\,1) in the visible range ($\lambda$\,$\leq$\,1\,$\mu$m). Due to the observing geometry of the Haumea (or any other TNO) ring, the actual phase angles are very low ($\alpha$\,$\lesssim$\,1\degr), and we can see backward-scattered light which is only a very small fraction of the total amount of scattered radiation, much smaller than in the case of isotropic scattering ($I(0)/I(\pi)$\,=\,1). 

\section{Results}
\label{sect:results}

\subsection{Spectral Energy Distributions of the Ring Models}

\begin{figure*}[ht!]
    \centering
    \hbox{\includegraphics[width=\columnwidth]{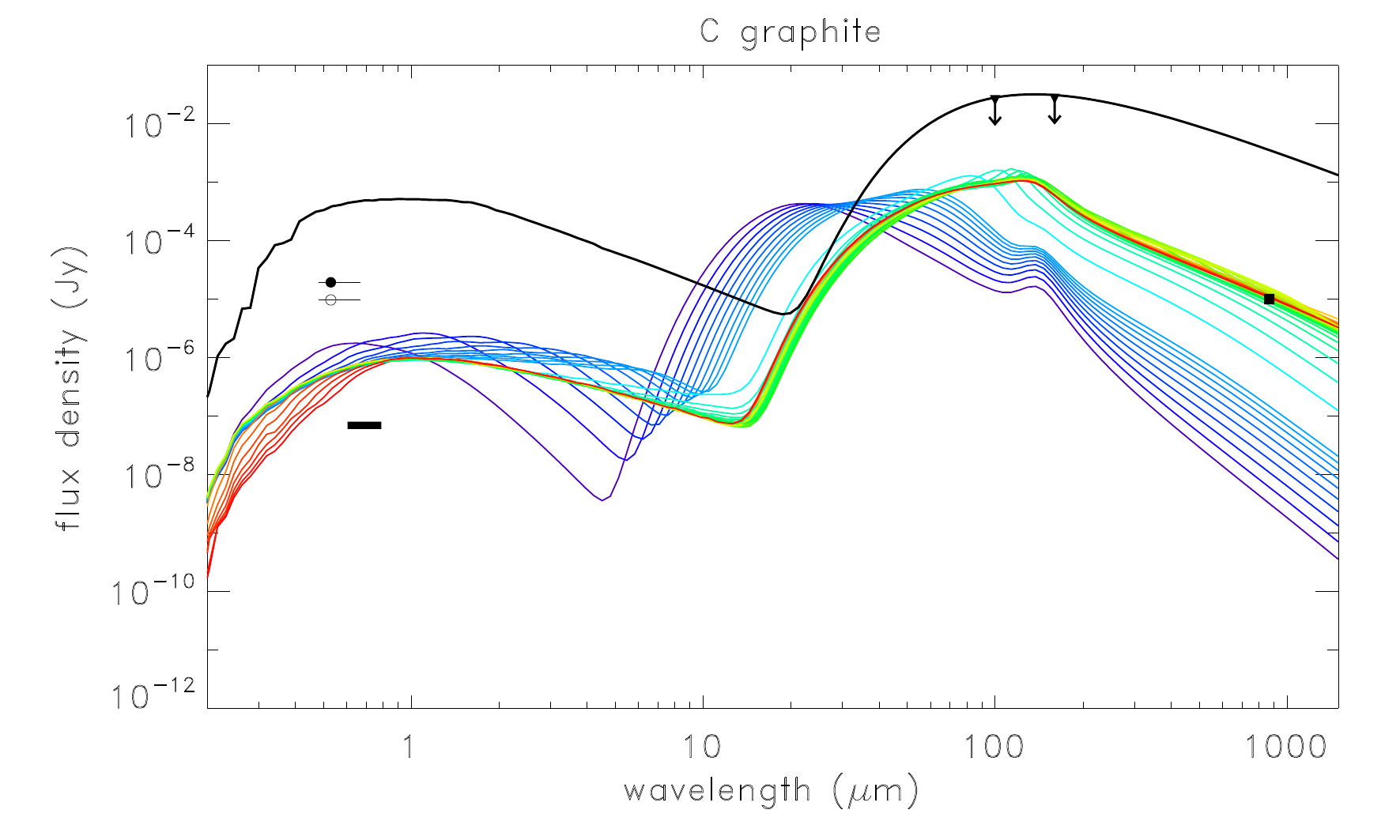}
    \includegraphics[width=0.16\columnwidth]{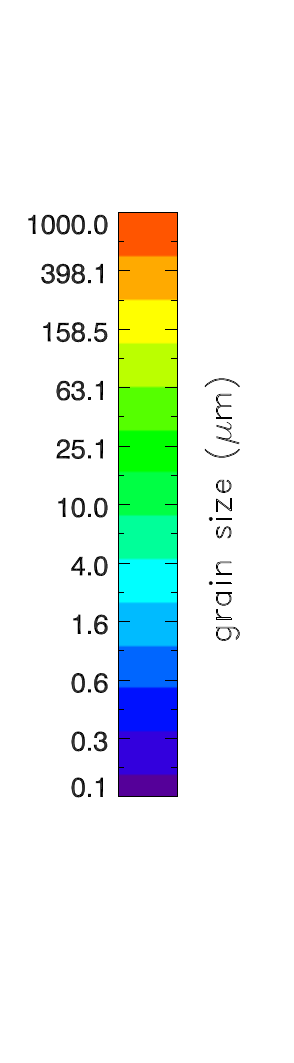}
        \includegraphics[width=\columnwidth]{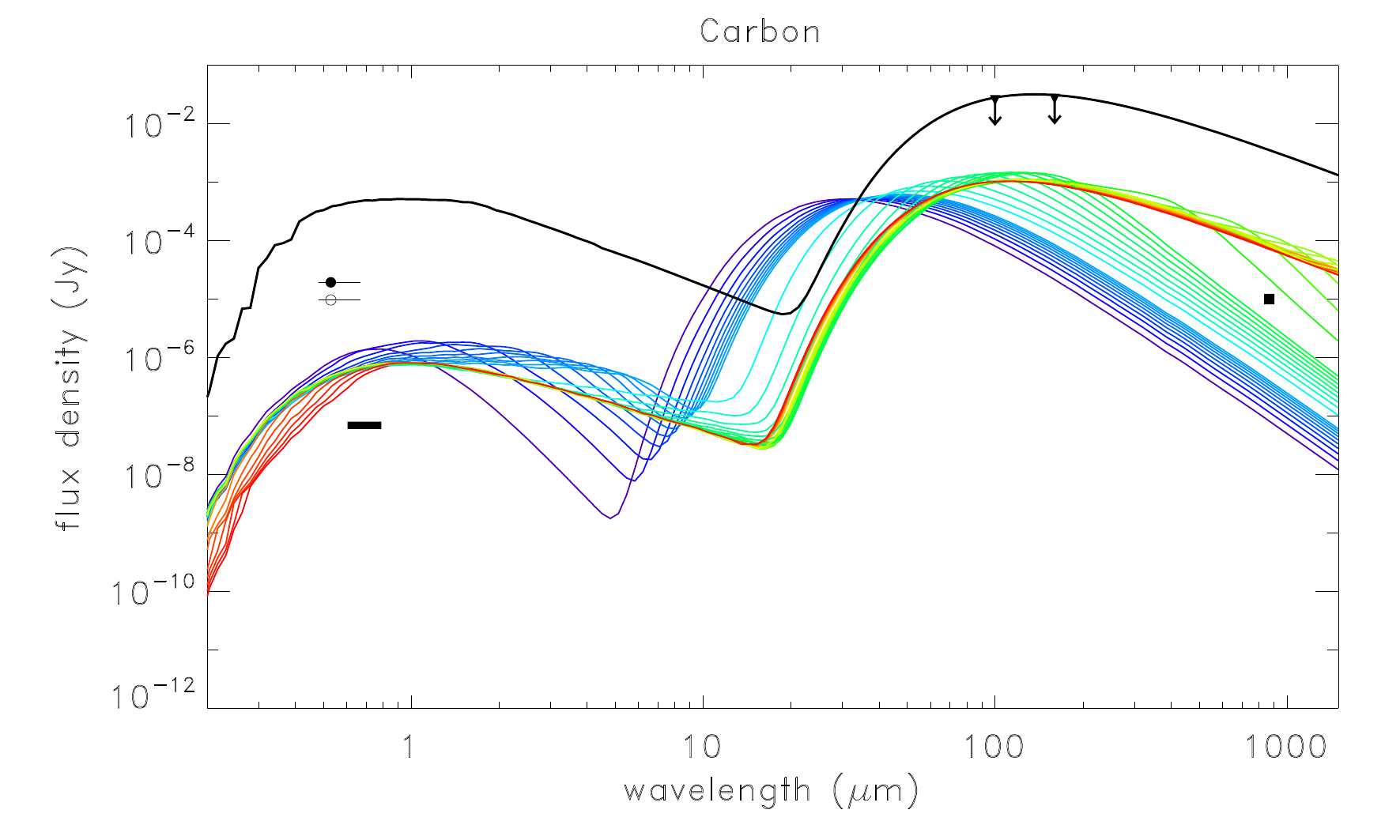}
        }
    \hbox{\includegraphics[width=\columnwidth]{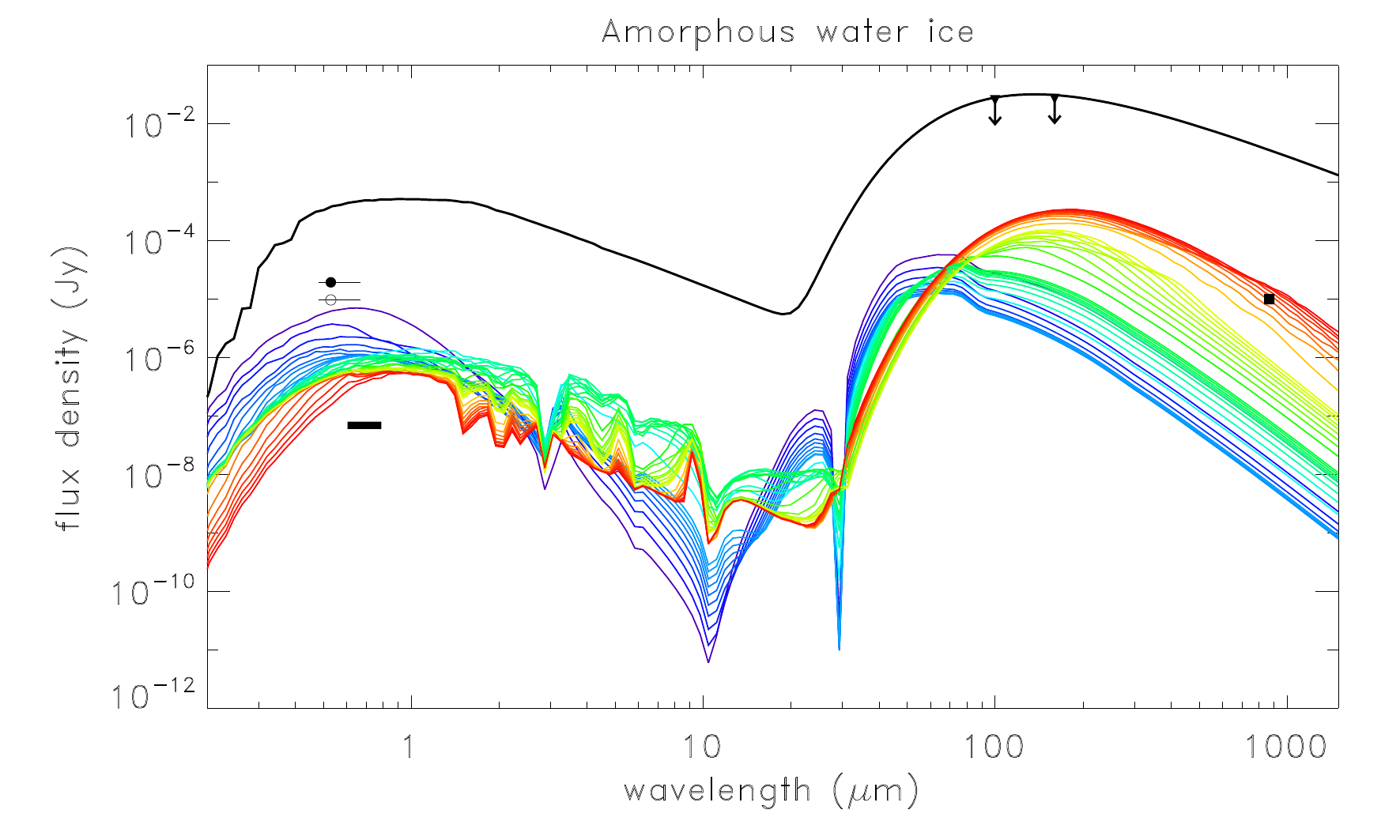}
    \includegraphics[width=0.16\columnwidth]{colorbar.pdf}
        \includegraphics[width=\columnwidth]{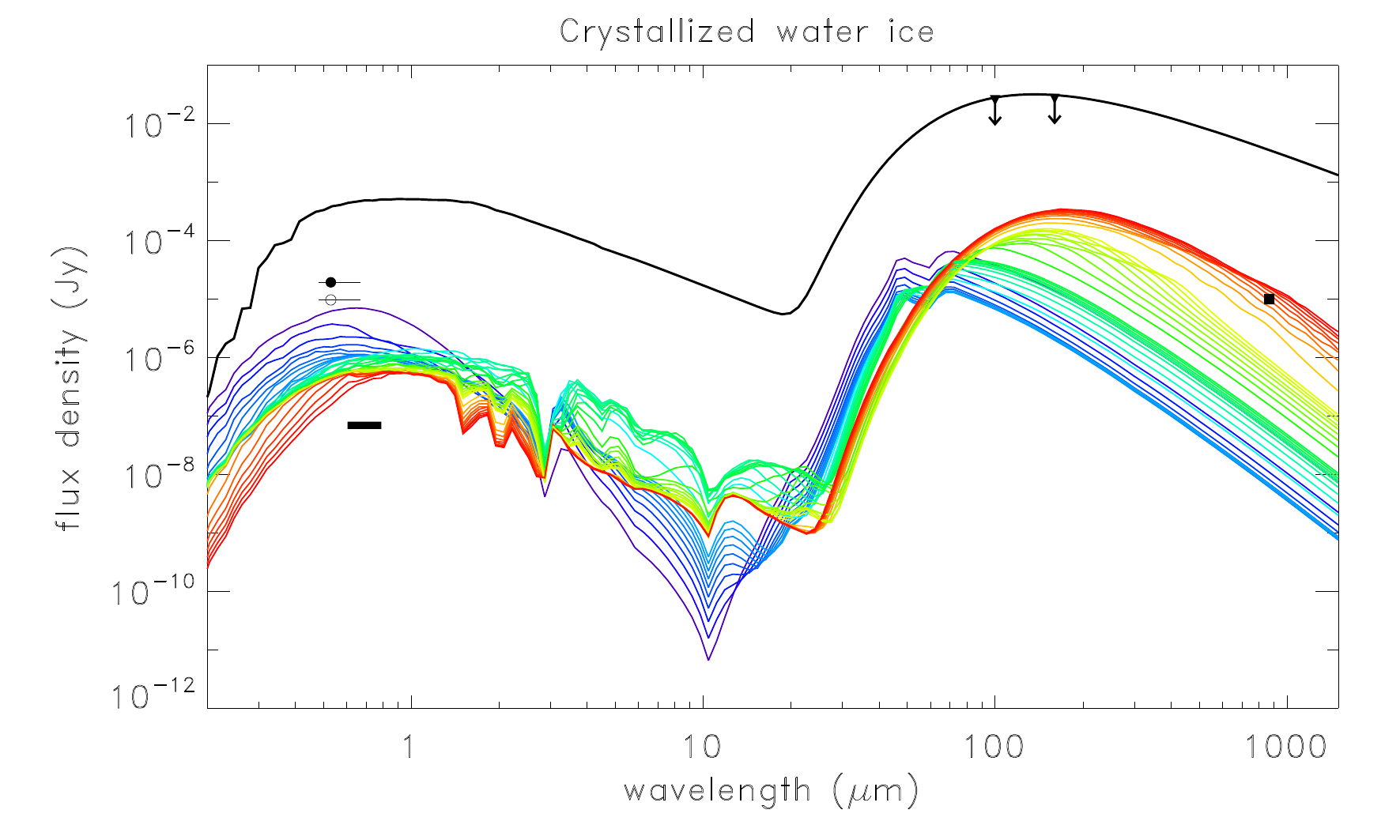}
    }
    \hbox{\includegraphics[width=\columnwidth]{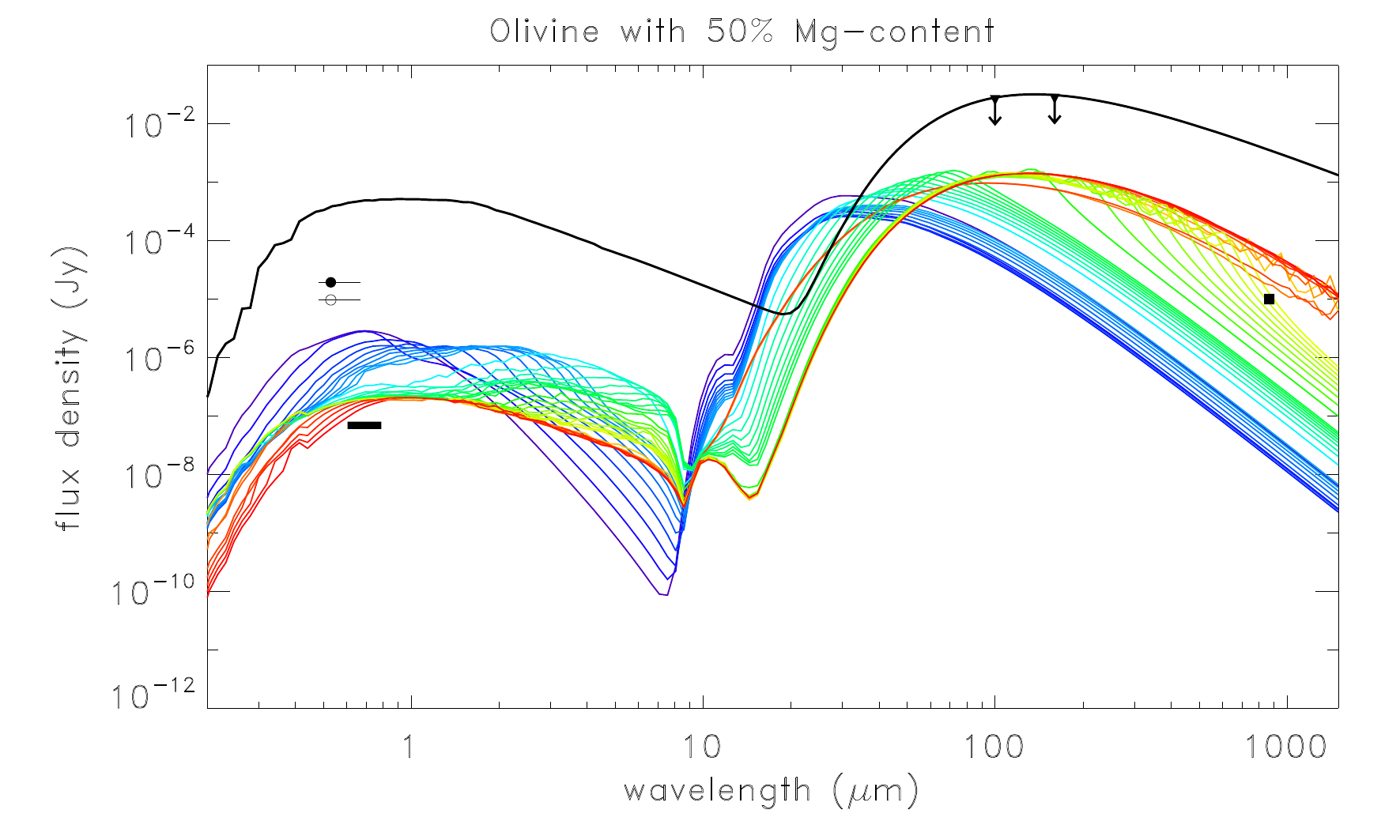}
    \includegraphics[width=0.16\columnwidth]{colorbar.pdf}
        \includegraphics[width=\columnwidth]{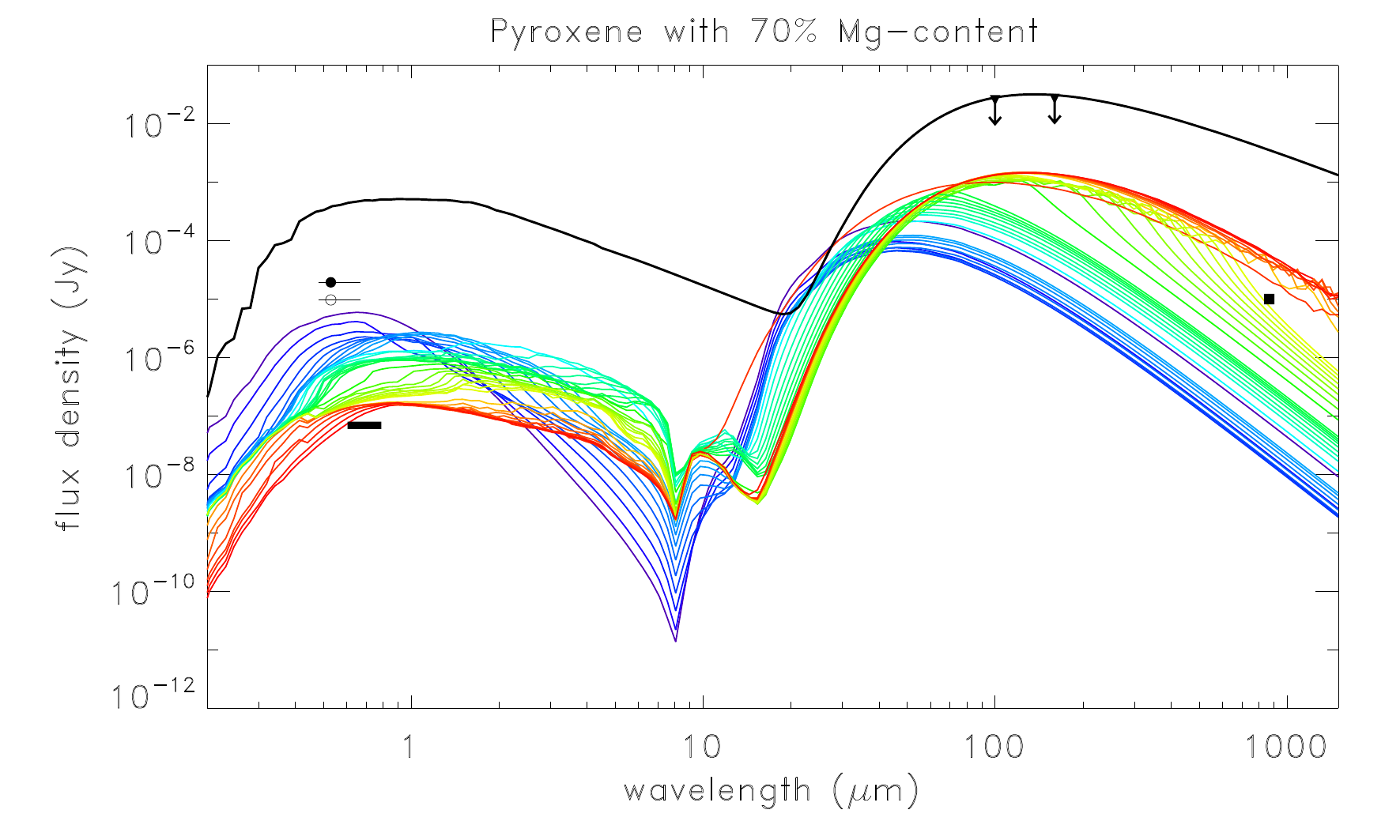}
    }
    \caption{Spectral energy distributions of the carbon graphite, amorphous carbon, amorphous and crystallized water ices, olivine (consists of 50\% magnesium) and pyroxene (composed of 70\% magnesium) models, calculated for the observing geometry of 2017, during the discovery of the ring system \citep{Ortiz2017}. All the models span grain sizes from 100\,nm to 1\,mm, indicated with different colours. The approximate SED of Haumea is displayed as a black curve. The filled circle represent the 5\% upper limit contribution of the ring, while the empty circle shows the $2.5\%$ best fit contribution from \cite{Ortiz2017} in Johnson $V$ band. Both circles are located at peak transmission, and the horizontal lines through them mark the wavelength range of the filter. The inverted triangles with the downward arrows represent the upper limits of the total brightness of the Haumea system at $100\,\mu$m and $160\,\mu$m based on \cite{Muller2019}. The black square represents a possible ALMA observation limit in band-7 ($\sim$870$\,\mu$m) with the antenna configuration C-9, while the black rectangle displays the wavelength range of the JWST NIRCAM F070W filter at a $5\sigma$ (point source) detection limit of 0.07 $\mu$Jy, with 1 hour integration. For more details, see Section 5.2.}
    \label{fig:seds}
\end{figure*}

In Figure \ref{fig:seds}, we show the SEDs of the graphite, carbon, water ice, olivine and pyroxene models. The different grain sizes are indicated with different colors. All the SEDs display a reflected light ($\lesssim 20\,$$\mu$m) and a thermal emission ($\gtrsim 20\,$$\mu$m) component. We marked the approximate SED of Haumea as a black curve, which was calculated using a Near-Earth Asteroid Thermal Model \citep[NEATM,][]{Harris1998}, assuming a volume-equivalent diameter of $D_{eq}$\,=\,1595\,km \citep{Ortiz2012,Muller2019}, and the latest geometric albedo and phase angle values, $p_V$\,=\,0.75 and q\,=\,0.44, derived by \citet{V22}. (We note that when using parameters from \cite{Ortiz2017} and matching a phase integral for $p_V=0.51$ \citep{V22} to recalculate the SED and to resolve the contradiction of the other geometric albedo with the occultation in 2017, the resulting SEDs almost perfectly overlap each other.) We used a beaming parameter of $\eta$\,=\,2.5 to match the observed far-infrared flux densities presented in \cite{Muller2019}. 

We also marked the observational constrains in the $V$ band, as determined by \cite{Ortiz2017}. The filled circle represents the 5\% upper limit contribution of the ring based on the rate of change of the amplitude of the rotational light curve. The empty circle shows the $2.5\%$ contribution, which was the best fit for the photometric data between 2005 and 2017. Both circles are located at the peak of the Johnson $V$ filter, while the horizontal lines through them indicate the wavelength range of the passband. The inverted triangles with the downward arrows represent the upper limits of the total brightness of the Haumea system at $100\,\mu$m and $160\,\mu$m \citep{Muller2019}. We also present the achievable observational limits of the current instruments that could separate the ring from Haumea itself. The black square represents the 10\,$\mu$Jy\,beam$^{-1}$ sensitivity of ALMA observation in band-7 ($\sim$345\,GHz, $\sim$870$\,\mu$m) with 8\,h integration at the antenna configuration C-9, 
and assuming third quartile weather conditions\footnote{see the ALMA Sensitivity Calculator: \url{https://almascience.eso.org/proposing/sensitivity-calculator}}. 
The black rectangle displays the wavelength range of the JWST NIRCAM F070W filter at a $5\sigma$ (point source) detection limit of 0.07 $\mu$Jy with 1 hour integration, as obtained with the JWST Exposure Time Calculator\footnote{\url{https://jwst.etc.stsci.edu/}}

\begin{figure}[ht!]
    \centering
    \includegraphics[width=\columnwidth]{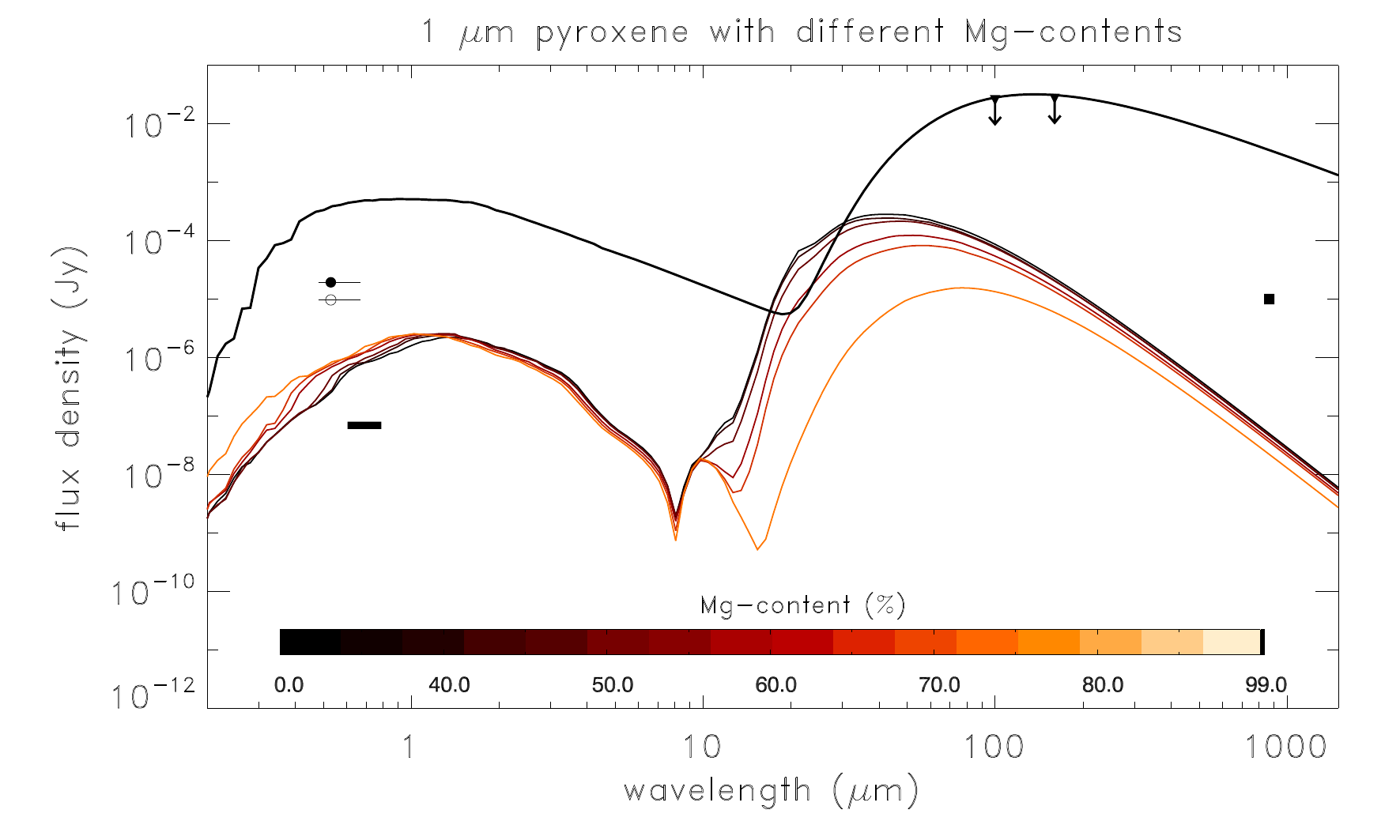}
    \includegraphics[width=\columnwidth]{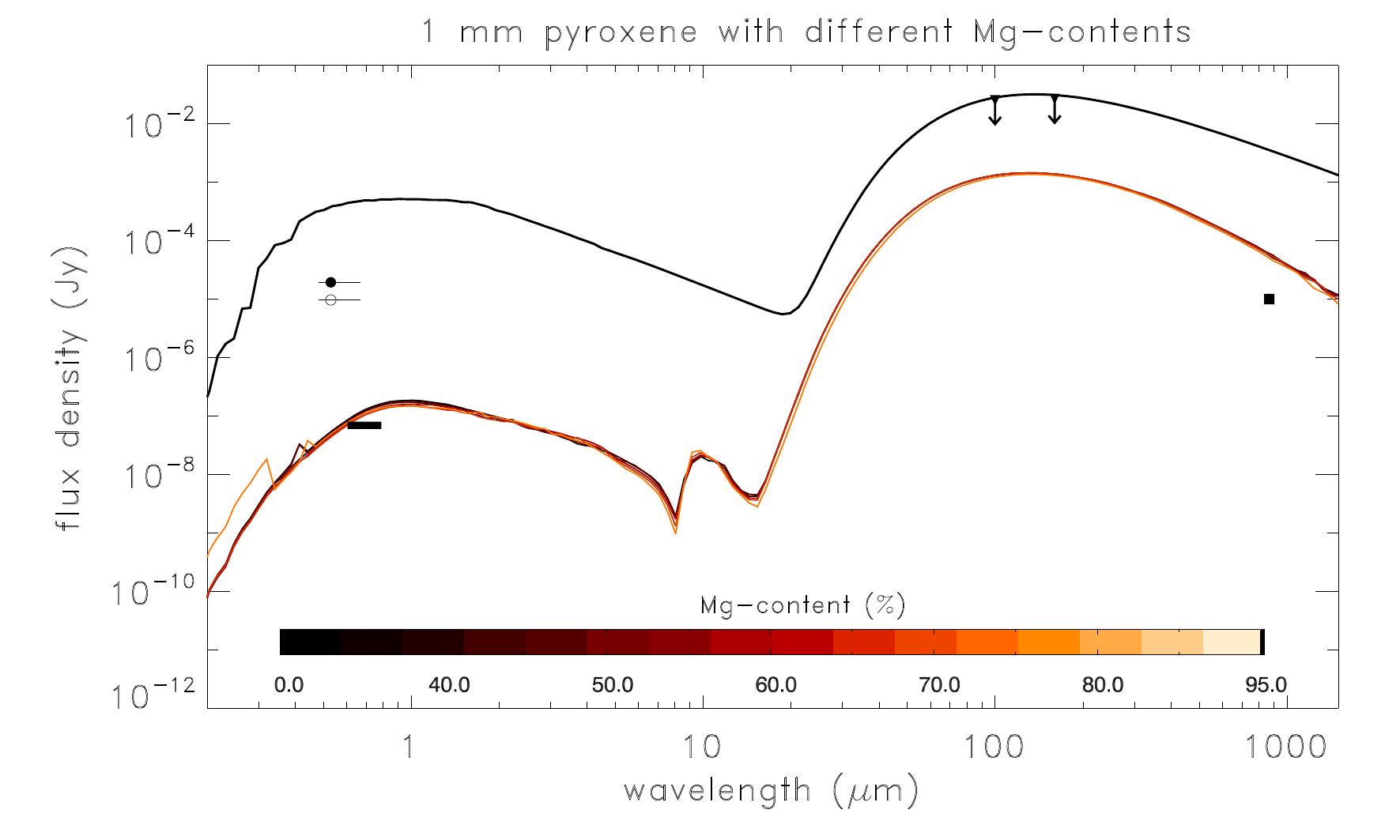}    
    \caption{The {\sl optool} package provides several built-in options for silicates containing different amounts of magnesium to calculate the absorption and scattering coefficients. Here we investigated the effects of the varying magnesium content using pyroxene models. We found that prominent differences can be seen only at the thermal emission part of the SEDs, which gradually decreases as the grain size increases. After reaching at particle radius of 1\,mm, all the pyroxene models tend to show the same SED.}
    \label{fig:comps}
\end{figure}

We note that {\sl optool} offers both amorphous and crystallized options for most materials, but we did not find significant differences at wavelengths of past and possible future observational facilities (see e.g., amorphous and crystallized water ices in Figure \ref{fig:seds}). Therefore, we decided to use the amorphous options in our analysis. We also compared different kinds of silicates (olivine or pyroxene) employing models with varying magnesium content. We found notable differences only at the thermal emission part of the SED, using particles with radii below 1\,mm (see an example for pyroxene models in Figure \ref{fig:comps}), while approaching the particle radius of 1 mm, all the models tend to show the same SED for a given material. 

\subsection{Comparison with Observational Constraints}

We calculated the $V$ band flux for all the models taking the $F(\lambda)$ SEDs of the models and the $V(\lambda)$ transmission curve of the Johnson $V$ filter into account:
\begin{equation}
    F_V = \frac{\int F(\lambda)V(\lambda)\lambda d\lambda}{\int V(\lambda)\lambda d\lambda}
\end{equation}
We found that the smaller the grain size, the larger the contribution of each ring in this passband, and all the models are below the $2.5\%$ best fit model of \cite{Ortiz2017}. The models closest to this estimation are the $100\,$nm water ice and the $100\,$nm pyroxene models with a contribution of $1.6\%$. For the extracted $F_V$ flux densities and their $F_V/F_{tot}$ brightness ratios for various compositions, see Table \ref{tab:v_band_ratio}.

\begin{table*}
    \centering
    \caption{Emission of the ring models with respect to the total brightness of the Haumea system for different materials such as graphite, amourphous carbon, water ice, olivine (with Mg-50\%) and pyroxene (with Mg-70\%). All the models were calculated for the 2017 observing geometry of the system. We calculated $F_V$ using Eq.~4. and exhibit our results here for several grain sizes. We take the total flux of the Haumea system as $F_{tot}=3.869\cdot 10^{-4}\,$Jy. In the brackets, we display the ring contribution to the total flux of the system. We found that the greatest contributions always comes from the smallest particles, and all the models are below the $2.5\%$ best fit model of \cite{Ortiz2017}. The models closest to the \cite{Ortiz2017} estimation are the $100\,$nm water ice and the $100\,$nm pyroxene models with an estimated contribution of $1.6\%$.}
    \label{tab:v_band_ratio}
    \begin{tabular}{cccccc}
    \hline
    \hline
         material & $F_V$ (Jy) for 0.1\,$\mu$m & $F_V$ (Jy) for 1\,$\mu$m & $F_V$ (Jy) for 10\,$\mu$m & $F_V$ (Jy) for 100\,$\mu$m & $F_V$ (Jy) for 1000\,$\mu$m\\
         \hline
          graphite & 1.702$\cdot 10^{-6}$ (0.440\%) & 8.626$\cdot 10^{-7}$ (0.223\%) & 8.774$\cdot 10^{-7}$ (0.227\%) & 9.682$\cdot 10^{-7}$ (0.250\%) & 7.505$\cdot 10^{-7}$ (0.194\%) \\
          carbon & 1.546$\cdot 10^{-6}$ (0.400\%) & 7.600$\cdot 10^{-7}$ (0.196\%) & 7.833$\cdot 10^{-7}$ (0.202\%) & 8.403$\cdot 10^{-7}$ (0.217\%) & 7.616$\cdot 10^{-7}$ (0.197\%) \\
          water ice & 6.380$\cdot 10^{-6}$ (1.649\%) & 1.163$\cdot 10^{-6}$ (0.301\%) & 1.251$\cdot 10^{-6}$ (0.323\%) & 9.108$\cdot 10^{-7}$ (0.235\%) & 5.540$\cdot 10^{-7}$ (0.143\%) \\
          olivine & 3.116$\cdot 10^{-6}$ (0.805\%) & 4.191$\cdot 10^{-7}$ (0.108\%) & 2.169$\cdot 10^{-7}$ (0.056\%) & 2.173$\cdot 10^{-7}$ (0.056\%) & 1.903$\cdot 10^{-7}$ (0.049\%) \\
          pyroxene & 6.015$\cdot 10^{-6}$ (1.578\%) & 1.292$\cdot 10^{-6}$ (0.334\%) & 9.610$\cdot 10^{-7}$ (0.248\%) & 2.037$\cdot 10^{-7}$ (0.053\%) & 1.478$\cdot 10^{-7}$ (0.038\%) \\
          \hline
    \end{tabular}
\end{table*}

\cite{Muller2019} estimated upper limits for the brightness of the whole Haumea system at 100\,micron and 160\,micron to be $2.52\cdot 10^{-3}$\,Jy and $2.66\cdot 10^{-3}$\,Jy, respectively. Our sum of the contribution of the main body and its ring aligns well with these constraints at the far-infrared. However, the mid-infrared region of carbon-rich and silicate models with lower magnesium content ($\lesssim$\,70\%) seem to offer a unique opportunity to directly measure the ring contribution, even if the system cannot be resolved. The corresponding models in Figure \ref{fig:seds} show that between 10-30 micron, the SED of the ring can exceed the thermal emission of Haumea itself. This mid-infrared excess of the SED can be a tracer of smaller (0.1--10\,micron) dust grains around the dwarf planet, as detected recently in the mid-infrared measurements of Makemake \citep{Kiss2024b}.

\begin{figure}[ht!]
    \centering
    \includegraphics[width=\columnwidth]{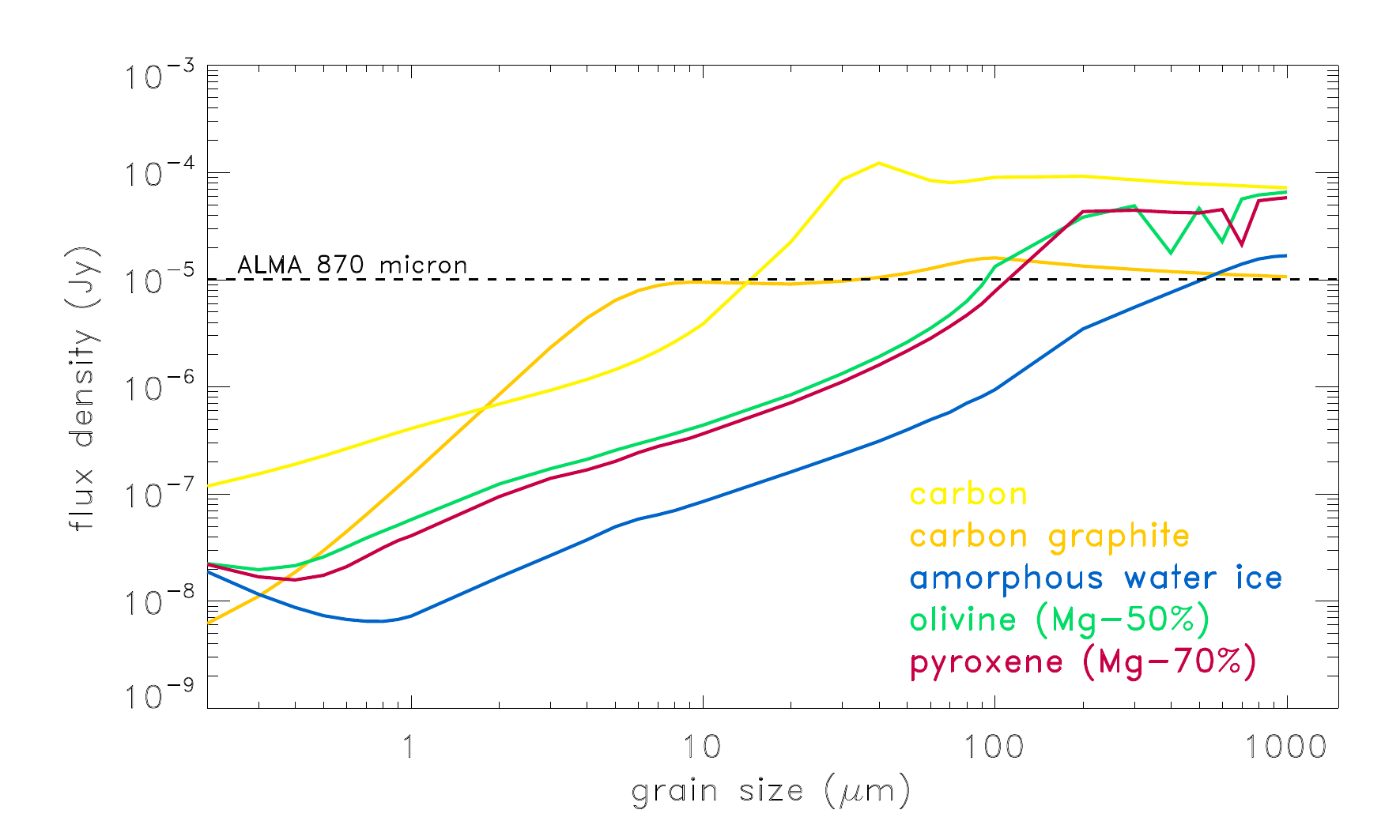}
    \includegraphics[width=\columnwidth]{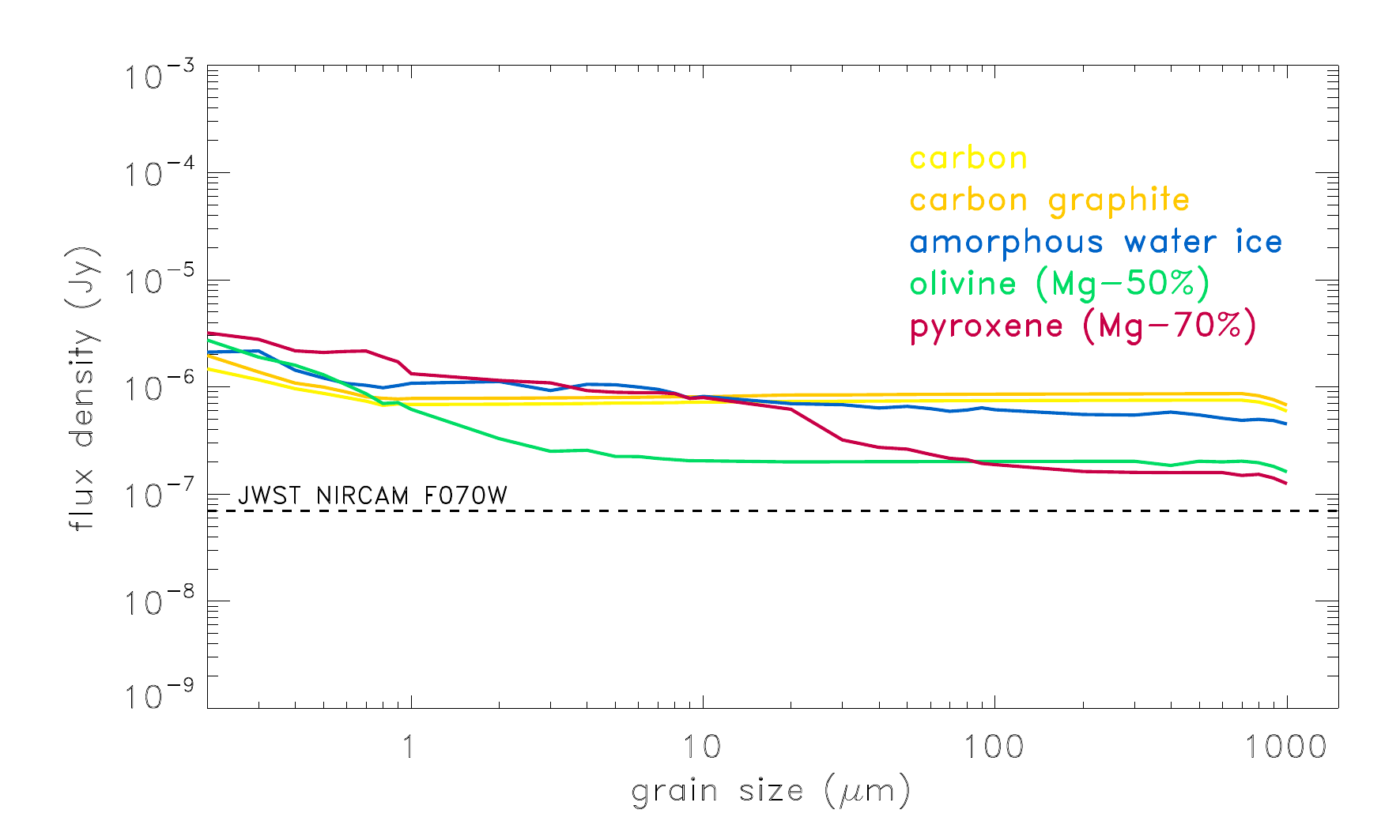}
    \caption{Detectability of the Haumea ring system. Flux densities have been derived at $\sim$\,870\,micron for ALMA (upper panel) and at the NIRCAM F070W passband for JWST (lower panel). We used our models calculated for the 2017 observing geometry for every grain sizes of amorphous carbon, graphite, amorphous water ice, olivine (with a magnesium content of 50\%) and pyroxene (with a magnesium content of 70\%) models. Combining ALMA and JWST to sample the reflected light and the thermal emission part of the SED respectively enables us to constrain possible ring compositions.}
    \label{fig:2017}
\end{figure}

Examining the rest of the energy distribution such as the visible or submillimeter wavelengths, we find that detection requires ring-only flux measurements. Currently, only the ALMA radio telescope and the James Webb Space Telescope can provide the sufficient spatial resolution to properly separate the ring from the main body, and the sensitivity to detect the ring as distinguished from the (much) brighter primary. In Figure \ref{fig:2017}, we investigate the detectability of the ring with ALMA and JWST based on our models calculated for the 2017 observing geometry of the Haumea system. We extracted the flux densities at $\sim$\,870\,micron for every grain size of the carbon, water and silicate models. In the upper panel of Figure \ref{fig:2017}, we can see that carbon-rich models can be observed from $\gtrsim$\,10\,$\mu$m with ALMA, while silicate models at around $\gtrsim$\,100\,$\mu$m, and pure water ice rings only in the millimeter regime. Here, we note again that using the simple ring model described in Section \ref{sect:simpleringmodel}, we got a single $F_{\mathrm{ALMA}}^{2017}=181\,\mu$Jy flux prediction for 2017. This example clearly shows the versatile capabilities of the radiative transfer approach over the simple ring model, which only includes the absorbed solar radiation of the Sun for the ring particles, and neglects the effect of thermal emission or the reflected sunlight. 

In the lower panel of Figure \ref{fig:2017}, we obtained the flux densities for JWST by convolving the NIRCAM F070W filter profile in a similar way as we performed for the Johnson $V$ filter using Eq.~4. We can notice that all the different composition models can be detected with this instrument. In this case, the flux densities over the grain sizes do not change as dramatically as in the submillimeter range, but still vary over one and a half order or magnitude.

\begin{figure}[ht!]
    \centering
    \includegraphics[width=\columnwidth]{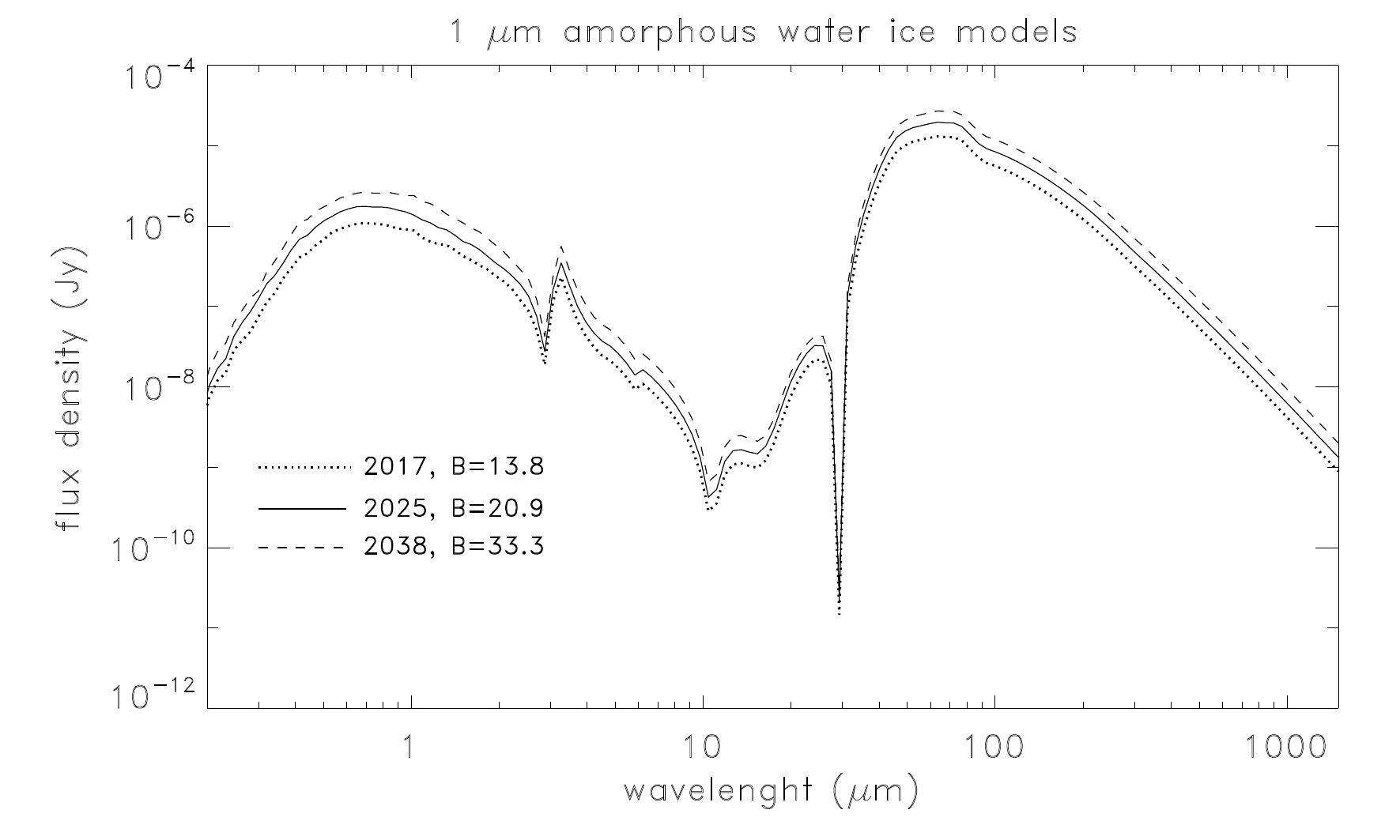}
    \caption{Since 2005, the opening angle ($B$) of Haumea's ring has been increasing. Here we show amorphous water ice models with a grain size of $1\,\mu$m using different opening angles. During the discovery of the ring was $B=13.8^{\circ}$ \citep{Ortiz2017}, and it is rapidly changing since then. In 2025, it will be $B=20.9^{\circ}$, while in 2038, $B=33.3^{\circ}$. In this figure, we show the difference between the respective spectral energy distributions to visualize the rising emission.}
    \label{fig:change}
\end{figure}

During the discovery of the ring in 2017, its opening angle was $B=13.8^{\circ}$ \citep{Ortiz2017}. Due to the orbital motion of Haumea, this angle is increasing since 2003, and it will rapidly change between 2025 and 2040 from $B=20.9^{\circ}$ to $B=35.2^{\circ}$. This means that since the discovery, the ring has been brightening significantly (see an example in Figure \ref{fig:change}), providing us improving conditions to observe the ring itself.

\section{Summary and conclusions}
\label{sect:summary}

During the last decade, occultation measurements showed that not only giant planets can host a ring system: Chariklo, Haumea and Quaoar are also harbouring rings with a great variety of properties. Some similarities are already remarkable, such as all the rings are close to the 1:3 spin-orbit resonance, or the presence of multiple rings in a system. However, the formation, evolution and stability of these rings are still open questions, to which information on the ring material are indispensable. Characteristic properties of the known small body ring systems predominantly comes from occultation measurements, obtained almost exlcusively in the visible range, but such observations provide very limited opportunities.

In this paper, we examined whether the currently available instruments are capable of resolving known small body rings in order to detect the ring-only emission, separate from the main body. We highlight the current sole detectability of the Haumea ring among the known Centaur/TNO ring systems using both the ALMA radio telescope at $\sim 870\,\mu$m and the James Webb Space Telescope's NIRCam instrument in the F070W filter configuration. At the present, only these facilities enable direct sampling of the thermal regime and the visible scattered light of the spectral energy distribution of the ring, which can be used to constrain the ring properties for a small body for the first time. This possibility can make the Haumea system a case study for any other small body systems.

We performed radiative transfer calculations using various materials and grain sizes to construct SEDs of different composition models. This is the first time that a radiative transfer approach was applied to derive flux densities for a small body's ring in the literature, overcoming the limitations of the previously extensively used simple ring models. We computed the spectral energy distributions using the opening angle of the ring from 2017 to obtain the contributions of the ring to the total brightness of the system in $V$ band estimated by \cite{Ortiz2017} in the discovery paper of the ring. We found that none of our models reach their $2.5\%$ best fit contribution, and the closest matches with $1.6\%$ are given by the smallest, 100--200\,nm water ice and pyroxene (with 70\% magnesium content) grains. However, most of the models show a brightness ratio of around 0.1--0.2\%. Comparing the SED of the ring models to that of the main body, we report a possibly observable mid-infrared excesses around 10--30\,micron, where carbon-rich or silicate models, consisting of a lower amount of magnesium can be brighter than the thermal emission of Haumea itself. We propose that this mid-infrared excess of the SED can be a tracer of smaller (0.1–10 micron) dust grains around any small body system, as we can see in the case of Makemake \citep{Kiss2024b}. We also note that in this analysis, we investigated small grains in order to accurately characterize their ring emission properties. While most of the rings around giant planets are made up of metre-sized boulders, there are also a few rings that consist of small particles, although these rings require special conditions, as they are much easier to remove, e.g. by radiation pressure. Collision of boulders, or intermittent release of material from the surface can generate/replenish finer grains. Our analysis also provides a first step for further investigations on rings with a mixture of large boulders and small grains, and/or rings created by using certain size distributions. 

We show that due to the increasing opening angle of the system, the ring emission will gradually rise in the coming years and decades. Sampling the ring-only flux in this way and comparing it with our models can provide direct constraints on the ring properties. Such information on the possible materials and grain sizes of the ring are crucial to develop formation theories.

\section*{Acknowledgements}
This research was supported by the `SeismoLab' KKP-137523 \'Elvonal, K-138962 and TKP2021-NKTA-64 grants of the Hungarian Research, Development and Innovation Office (NKFIH). Cs. Kalup was supported by the \'UNKP-23-3 and \'UNKP-24-3 New National Excellence Program of the Ministry of Culture and Innovation from the source of the National Research, Development and Innovation Fund. This research made use of NASA’s Astrophysics Data System Bibliographic Services.





\bibliography{ref}{}
\bibliographystyle{aasjournal}


\end{document}